**A technique to detect periodic and non-periodic ultra-rapid flux time variations with standard radio-astronomical data**

Ermanno F. Borra[1*], Jonathan D. Romney[2] and Eric Trottier[1]

[1] Département de Physique, Université Laval, Québec, Qc, Canada G1V 0A6

[2] National Radio Astronomy Observatory, Socorro, New Mexico, USA 87801

[*] email: borra@phy.ulaval.ca

Address: Département de Physique, Université Laval, Québec, Canada G1V 0A6

Email : borra@phy.ulaval.ca

**ABSTRACT**

We demonstrate that extremely rapid and weak periodic and non-periodic signals can easily be detected by using the autocorrelation of intensity as a function of time. We use standard radio-astronomical observations that have artificial periodic and non-periodic signals generated by the electronics of terrestrial origin. The autocorrelation detects weak signals that have small amplitudes because it averages over long integration times. Another advantage is that it allows a direct visualization of the shape of the signals, while it is difficult to see the shape with a Fourier transform. Although Fourier transforms can also detect periodic signals, a novelty of this work is that we demonstrate another major advantage of the autocorrelation, that it can detect non-periodic signals while the Fourier transform cannot. Another major novelty of our work is that we use electric fields taken in a standard format with standard instrumentation at a radio observatory and therefore no specialized instrumentation is needed. Because the electric fields are sampled every 15.625 nanoseconds, they therefore allow detection of very rapid time variations. Notwithstanding the long integration times, the autocorrelation detects very rapid intensity variations as a function of time. The autocorrelation could also detect messages from Extraterrestrial Intelligence as non-periodic signals.

1. Introduction

   Astronomical objects that have rapid periodic flux time variations are known to exist (e.g. pulsars). The most rapid flux variations in millisecond pulsars can have periods as short as a 1.4 milliseconds (Hessels et al. 2006). Measurements of these flux variations in pulsars are done with specialized instrumentation. It is possible that some exotic objects, like quasars or Active Galactic Nuclei (AGNs) may emit periodic pulses that have very short periods. This hypothesis is validated by the discovery of periodic modulations in the spectra of galaxies (Borra 2013). These periodic modulations could be caused by the emission of pulses separated by times shorter than $10^{-10}$ seconds (Borra 2010). Finding these periodic pulses with conventional techniques would be difficult since the periods could be extremely short and most quasars and AGNs are distant faint objects. Furthermore expensive specialized instrumentation would be needed. It is also possible that some exotic objects emit rapid pulses that are not periodic, like the pulses of Fast Radio Bursts. They would be very difficult to detect with conventional techniques. The pulses of Fast Radio Bursts last a few milliseconds (e.g. FRB 110523 in Masui et al. 2015 and FRB121102 in Spitler et al. 2016) and are easily detected with conventional techniques, while the pulses that we are considering in this paper occur over time scales that are much shorter (less than $10^{-10}$ seconds).

   We shall consider a very simple numerical technique that uses the autocorrelation of intensity variations as a function of time $I(t)$ to detect rapid periodic and non-periodic intensity variations. The advantage of the technique is that it can easily detect very rapid intensity time variations as well as intensity variations that have very small amplitudes and are buried in noise in $I(t)$. This is because the autocorrelation can average over a long time, thereby allowing the detection of very weak signals, as well as intensity variations that occur over times that are only moderately longer than the instrumental sampling time. Detecting a signal over long integration times requires that the signal be present over a substantial fraction of the time, otherwise it would be buried in noise. Red noise (also called Brown noise) is commonly present in radio astronomical data. The characteristic of red noise is that it varies inversely proportional to the square of the frequency. It will have no significant impact on our use of the autocorrelation, because we autocorrelate data taken within very small frequency channels, each 32 MHz wide. The signals that we detected were obviously not obliterated by red noise. We use data that are obtained with existing radio interferometers in the standard format used for interferometric observations. The major advantage of using standard interferometric radio-astronomical data is that one does not need specialized instrumentation and that, because the samplings are done over very short times, one can detect very small flux variations. The software needed for the autocorrelation is present in commercially available software (e.g. Mathematica and Matlab) and, if needed, is very easy to write since one only needs to perform a numerical integration (see Equation 1 in the next section). The software that we used was written by us.

Using the autocorrelation to find periodic variations is a known technique; however, a novelty of our work is that we demonstrate it using standard radio-astronomical data that are sampled every 15.625 nanoseconds, thereby allowing the detection of very rapid time variations. The major novelty of our work is that we will demonstrate that the autocorrelation can detect non-periodic variations. As far as we know, the autocorrelation has never been used for finding non-periodic variations at nano-second time-scales with astronomical data.

Other techniques can also be used to find periodic signals (Ransom et al. 2002) as well as transients (Law & Bower 2012) in digital data. We compare our technique to other techniques in the discussion, where we illustrate the advantages of our technique.

2. Demonstration of the use of the autocorrelation technique with radio-astronomical data

To detect intensity time variations we use the autocorrelation of the intensity as a function of time *I(t)* to which we subtract the average of the intensity over time <*I(t)*>, where the brackets < > signify a time average of *I(t)* which is carried out over the entire time of observation. The autocorrelation at a time lag $\tau$ is therefore done using $\Delta I(t) = I(t) - <I(t)>$

$$R(\tau) = \int_0^T \Delta I(t+\tau) \Delta I(t) dt \quad , \quad (1)$$

where *T* is the entire time of observation of *I(t)*. The reason we carry out this subtraction is that the autocorrelation of *I(t)* gives a *R(t)* with a very steep slope, while the autocorrelation of $\Delta I(t) = I(t) - <I(t)>$ gives a horizontal line with an average value near 0, making it easier to detect signals. The numerical integral in Equation 1 had to be done with a supercomputer because of the huge quantity of data needed due to the small time sampling (15.625 nanoseconds} that we used.

We use data observed by the Very Long Baseline Array (VLBA; Napier et al. 1994, Romney & Reid 2005), operated by the US National Radio Astronomy Observatory (NRAO). The data were taken in one of the VLBA's standard operating modes for cross-correlation interferometry, in 10-minute segments of test time in order to evaluate the new analysis technique described in this paper. A total of 16 dual-polarization channels, each 32 MHz wide, spanned frequencies covering a contiguous range of 8256-8512 MHz. Raw data were extracted from the standard recording media, and transmitted to us via an Internet connection. The data consist of an electric field *E(t)* that is sampled every 15.625 nanoseconds with a digital sampling of only 2 bits. These 2 bits can only give 4 binary values (00, 01, 10, 11) of the electric field *E(t)* that are converted to 0, 1, 2, 3 in decimal notation. We then use C++ software to convert these four decimal values to 4 electric field values (-3.3359, -1, 1, 3.3359). The details of this are discussed in section 8.3 of the book written by Thompson, Moran and Swenson (2004). The intensity as a function of time *I(t)* is computed from the time average of the square of the electric field, $I(t) = <E(t)^2>$. We carry out the time average over 90 samples (1.4 microseconds).

Some of the data contained man-made periodic signals that were totally invisible in plots of the intensity as a function of time *I(t)* but the autocorrelation easily detected these periodic signals.

We shall demonstrate the use of the autocorrelation in detecting periodic signals with standard VLBA calibration signals which were active as usual in an early experiment in this series. The first of these is a periodic rectangular signal with a period of 80 Hz. This is a switched-noise signal that is introduced in the data to measure the system noise power for calibration of the cross-correlation amplitude in interferometric measurements. Figure 1 shows plots of the intensity as a function of time for 2.5 million samples (39 milliseconds) of observations. Note that the rectangular signal in *I(t)* is totally invisible in the original intensity data which simply looks like noise. To render visible the rectangular periodic signal we had to carry out a smoothing using a moving average filter having a width of 90 samples (1.4 microseconds) to generate Figure 1. Figure 2 shows the autocorrelation done using Equation 1. We only used the same sample displayed in Figure 1 (39 milliseconds of observations) to generate Figure 2 with the autocorrelation. The signal that is barely detectable in Figure 1, even after the smoothing, gives a very visible signal in the autocorrelation, thereby clearly illustrating the advantage of using the autocorrelation. To generate Figures 1 and 2, we use a signal that can be seen in the smoothed data for demonstration purposes. A much weaker signal with the same period would not be visible with that smoothing. To see a signal in smoothed intensity data would require appropriate smoothing. Too long a smoothing would destroy a short period signal and too brief a smoothing would not allow seeing a weak long period signal, so that one would need an estimate of the strength and duration of the pulses to carry out an appropriate smoothing. A weak signal having a short period could not be seen after smoothing since the smoothing would have to be considerably longer than the period On the other hand, the autocorrelation would allow detection without the use of unknown parameters, carry out an average over a very long time and still detect very weak signals having very short period. For example, this is the case for the signals discussed in the next two paragraphs and shown in Figures 4 and 5.

The second VLBA calibration signal detected by the technique is a 5-MHz comb, produced by very narrow pulses injected on an exact 200 nanosecond period. Figure 3 shows only the intensity variations as function of instrumental sampling. The pulses seen in Figure 3 are caused by the instrumentation sampling time (15.625 nanoseconds). The 5 MHz injected pulses are totally invisible in plots of the intensity versus time of data obtained from observations of the quasar because they are hidden in noise. The signal is also invisible if one makes an autocorrelation that uses only the same number of samples of observations (2.5 million samples) that were used to generate Figure 1. However the calibration signal becomes highly visible if one makes the autocorrelation using a much longer autocorrelation time. Figure 4 shows the first 70 samples of an autocorrelation that uses 3.2 billion samples (50 seconds) of data. The periodic signal can clearly be seen in Figure 4. We plot the intensity as a function of sampling number to clearly show that the autocorrelation can detect time variations comparable to the sampling time.

The third demonstration of the technique uses observations of the quasar 3C 345. The data contains an artificial signal that comes from a background quasi-periodic signal generated either by a radar located near one of the VLBA stations used to observe 3C 345 or some internal cause. The signal is very nearly periodic with an average period of 8 milliseconds. Figure 5 shows 120 milliseconds of the autocorrelation. The autocorrelation uses 38.4 billion samples (600 seconds) of data. The signal is clearly visible in Figure 5. The signal is totally invisible in plots of intensity versus time, even after smoothing.

Finally, we show that the autocorrelation can also be used to detect intensity time variations that are not periodic, have small amplitudes and vary over short time scales. If *I(t)* contains very small non-periodic intensity fluctuations, $\Delta I(t) = I(t) - <I(t)>$ in Equation 1 will only contain these fluctuations and fluctuations caused by noise. The values of $\Delta I(t)$ can be negative or positive. The integral in Equation 1 is carried over the entire time of observation but done over $\Delta I(t+\tau)\Delta I(t)$. When $\tau = 0$, the integral is carried out over $\Delta I(t)\Delta I(t)$ and the entire time of observation. The negative and positive fluctuations will all give a positive contribution to the integral because it is carried over the product $\Delta I(t)\Delta I(t)$ and therefore both negative and positive fluctuations give a positive contribution. The fluctuations therefore add up in the integral and will give a large value of $R(\tau)$, significantly above the noise, if there are a very large number of random fluctuations. Consequently $R(\tau)$ has a maximum value at $\tau = 0$. As $\tau$ increases, the integral is carried over $\Delta I(t+\tau)\Delta I(t)$ so that the fluctuations are no longer at the same positions and, consequently, there are positive and negative contributions in the integral so that $R(\tau)$ decreases. Another effect of increasing $\tau$ is that the fluctuations contained at the smallest and larger values of $t$ in $I(t)$ no longer overlap. They no longer contribute in the integral and therefore the value of $R(\tau)$ decreases as $\tau$ increases. For $\tau = T$ there is no overlap at all and the integral gives $R(\tau) = 0$. Figure 6 gives an example of an autocorrelation of an $I(t)$ that contains many non-periodic signals. It uses the autocorrelation of 650 seconds of observations (4.16 $10^9$ samples) of the quasar PMN 0134-0931. The data contain a large number of artificial non-periodic signals. The autocorrelation shows the type of signal discussed in the previous sentences. The noisy area surrounding the line is caused by the noise that is present in the data. The autocorrelation goes to negative values from 200 seconds up to 500 seconds and then increases again because of the subtraction of the average intensity discussed at the beginning of section 2.

3. Discussion and conclusion

We have demonstrated that rapid periodic and non-periodic weak intensity time variations can be found using the autocorrelation of intensity as a function of time. Although it is well-known that the autocorrelation can be used to detect periodic signals, it has never been used to detect non-periodic signals. A major interest of our demonstration is that this was done with radio-astronomical data taken in the standard format used for interferometric observations and no specialized instrumentation is required. Very rapid variations can be detected because the sampling time of the data is only 15.625 nanoseconds.

The main conclusion of this work is that the autocorrelation can easily be used to detect periodic and non-periodic time variations of the order of the time between samples. This can be seen in Figure 4, where an artificial signal having a period of 13 times the sampling time of 15.625 nanoseconds, shown in Figure 3, can clearly be seen in the autocorrelation. However, in practice, the lower limit to the period will be set by a smoothing effect coming from the dispersion caused by the ionized regions along the ray path from the cosmic source. This effect comes from the fact that radio waves at different frequencies move at different speeds in the galactic and intergalactic media and consequently the electric fields *E(t)* that are at different frequencies inside a bandpass arrive at different times at the telescope. This widening effect lengthens the pulses and consequently decreases their amplitudes, thereby making them more difficult to detect. However this effect is less important in the autocorrelation, because of its averaging effect. It will take a widening significantly larger than the width of the pulses to make intensity time variations undetectable with the autocorrelation. This can be seen in Figure 4, where the signal is strongly detectable even after a numerical smoothing with a width of 90 samples, which is about 7 times the period of 13 samples. The periodic signal was still visible in an autocorrelation that used a numerical smoothing of 7000 samples (538 times the period). At redshifts of $z = 0.5$, $z = 1.0$, $z = 2.0$ and $z = 3.0$ the widening caused by the intergalactic medium are respectively of 195, 360, 635 and 865 microseconds. Consequently periods of less than a microsecond should be detectable at these redshifts because, as discussed in this paragraph, the autocorrelation can detect periodic time variations less than 500 times smaller than the smoothing time. This widening effect coming from the galactic and intergalactic media is frequency dependent and decreases with increasing frequency. Consequently observations obtained at higher frequencies should be able to detect smaller periodic variations. This widening effect would also be smaller if one uses channels that have narrower bandpasses, because the widening is due to the fact that different frequencies travel at different speeds in the galactic and intergalactic media. Our channel bandpasses have a width of 32 MHz. The autocorrelation could also be used to find rapidly varying objects within our galaxy or nearby galaxies. In that case the only widening effect would come from the galactic medium and be less important, thereby allowing finding even faster time variations. It also could be used to find and study pulsars.

An advantage of the use of the autocorrelation comes from its simplicity, because one does not need specialized instrumentation, like the instrumentation used to observe pulsars. One simply uses data obtained with existing radio telescopes, in their standard digital format. Reading about our use of artificial signals to validate our technique, one may be worried about false detections coming from these artificial signals. Let us first say that the artificial calibration signals, like the signals used to generate Figure 2 and 4, can easily be turned off during observations that one would need to detect time variations. They were turned off during most of our observations. Artificial signals coming from other artificial sources, like the signal used to generate Figure 5, can easily be excluded by using, like we did, simultaneous observations with several interferometric telescopes located at different distant locations. An artificial signal would only be present in a single telescope. Note that the VLBA telescopes are separated by very large distances so that even atmospheric signals would not be present in separate telescopes. For example, the artificial background signal shown in Figure 5 was only detected in a single telescope. Note that a very large number of observations, used for a different purpose, do not have artificial signals. In this paper we discuss observations with artificial signals, because they validate the technique.

Other techniques that use the Fourier transform like the techniques used in the PRESTO software (e.g. Ransom et al. 2002) can also be used to find periodic signals in digital data. The outstanding advantage of the autocorrelation is that it can detect non-periodic signals while the Fourier transform cannot. Another advantage is that the autocorrelation can easily allow visualizing the shape and time variations of the autocorrelated signals. This can be seen in Figures 2 to 5, where one can clearly see the shape of the time variations of the autocorrelation of the signals. While the shape of the autocorrelation is not the same as the shape of the basic time-dependent signals, it gives the general characteristics of the signals (e.g. periodicity, average time duration of the pulses). While the Fourier transform easily identifies the main period of pulses, it is very difficult to visualize the shape of the pulses from the Fourier spectrum. Another advantage is the simplicity of the autocorrelation software, which is present in many commercially available software (e.g. matlab) and very easy to write if needed, as can be seen by looking at Equation 1. Finally note that an important novelty of our work comes from the fact that we used standard radio-astronomical data.

The technique discussed in Borra (2010) and used by Borra (2013) can also be used to find rapid intensity time variations. However it cannot be used to find pulses separated by time shorter than $10^{-10}$ seconds in optical spectra. The advantage of our technique is that it can detect periodic time variations significantly shorter than $10^{-10}$ seconds as well as quasi-periodic and non-periodic time variations. While conventional matched filtering can easily detect non-periodic signals of a few milliseconds duration, like those in Fast Radio Bursts, the advantage of the autocorrelation is that it can detect non-periodic nano-second pulses and therefore give a unique use in the ultra-rapid time variation domain.

Fast Radio Bursts (FRBs) are a recent astronomical discovery. The transient, flickering radio sky is not well understood. Consequently, analyzing data with novel techniques may lead to the discovery of unexpected complex physical phenomena. The first FRB was discovered by Lorimer et al. (2007). Since then,33 objects with FRBs have been discovered (http://www.frbcat.org/). They are powerful bursts that last a few milliseconds. The majority of the detections consisted of a single pulse with the exception of FRB121102 (Spitler et al. 2016) which had 10 repeat bursts, with 6 bursts occurring within a 10 minutes interval and 3 bursts weeks apart. Although the autocorrelation would not have particular advantages for the detection of a single FRB, it would allow the detection of FRBs that repeat within the time of observation. The autocorrelation will not be advantageous over conventional matched-filtering techniques for detecting periodic repeating bursts, particularly if these conventional techniques use the Fourier transforms. However, an advantage comes from the fact that it allows to visualize the shape of repeating bursts while it would be very difficult to visualize the shape in the Fourier Domain. In practice, as mentioned earlier, the major advantage of the autocorrelation would be to detect non-repeating bursts (non-periodic signals) that occur over very short time scales, while the Fourier transform would not be able to detect them. FRB121102 (Spitler et al. 2016) had 6 bursts in 10 minutes interval, within the 650 seconds interval that generated Figure 6, which was generated by non-periodic signals. Although far fewer in numbers than the pulses that generated Figure 6, the 6 FRB121102 bursts would give a detectable signals because they are much more powerful. The origin of FRBs is totally unknown so that any hypothesis is possible. It is therefore possible that some astronomical objects emit a large number of very weak FRBs that are totally undetectable with standard techniques. These numerous weak FRBs would generate exactly the type of signal shown in Figure 6. Law & Bower (2012) discuss a technique to detect radio transients; however, it is very complex, while ours is very simple. It depends on interferometric results while ours does not. Furthermore, Law & Bower (2012) do not demonstrate the use of their technique to detect numerous weak non-periodic pulses, like we did. Finally, let us note that there have been speculations that FRBs may come from extraterrestrial intelligence (Scoles 2015, Lingam & Loeb 2017). If some extraterrestrial intelligence (ETI) sends messages encoded in a large number of radio pulses weaker than FRBs, the autocorrelation could detect them as non-periodic pulses coming from stars. This is, of course, highly speculative but also highly interesting if ETI signals are ever discovered. Note also that a very large number of hypotheses of the sources of FRBS have been made. They are discussed, with a large number of references in the Origin hypotheses section of the Fast Radio Burst Wikipedia web site.

ACKNOWLEDGMENTS

This research has been supported by the Natural Sciences and Engineering Research Council of Canada

The National Radio Astronomy Observatory is a facility of the US National Science Foundation operated under cooperative agreement by Associated Universities, Inc.

FIGURES

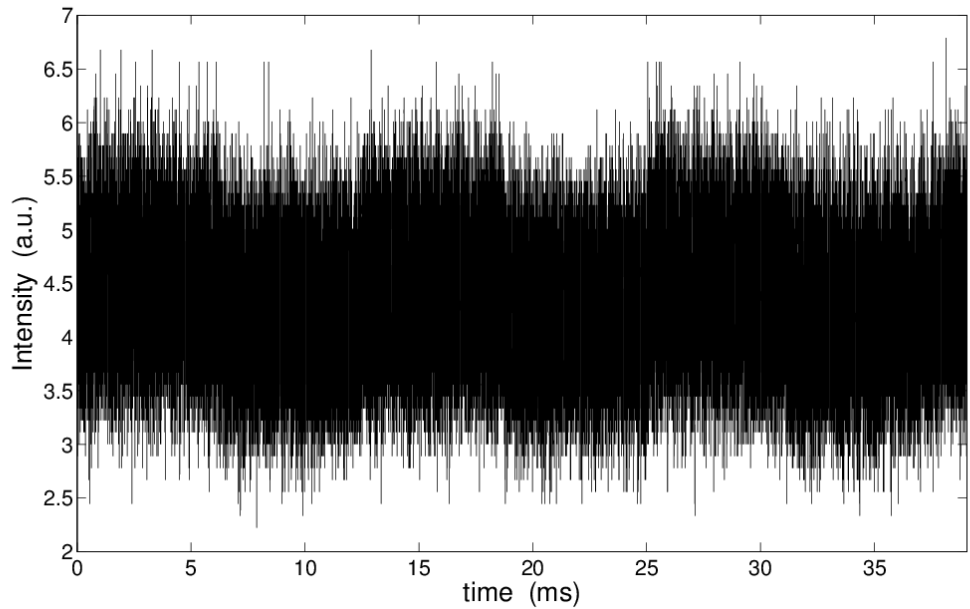

Figure 1

Intensity (in arbitrary units) as a function of time for 2.5 million samples (39 milliseconds) of observations of the quasar PKS 1622-253. A smoothing, using a moving average filter having a width of 90 samples (1.4 microseconds), had to be made to render the rectangular signal visible because it is totally invisible in plots that do not use a smoothing.

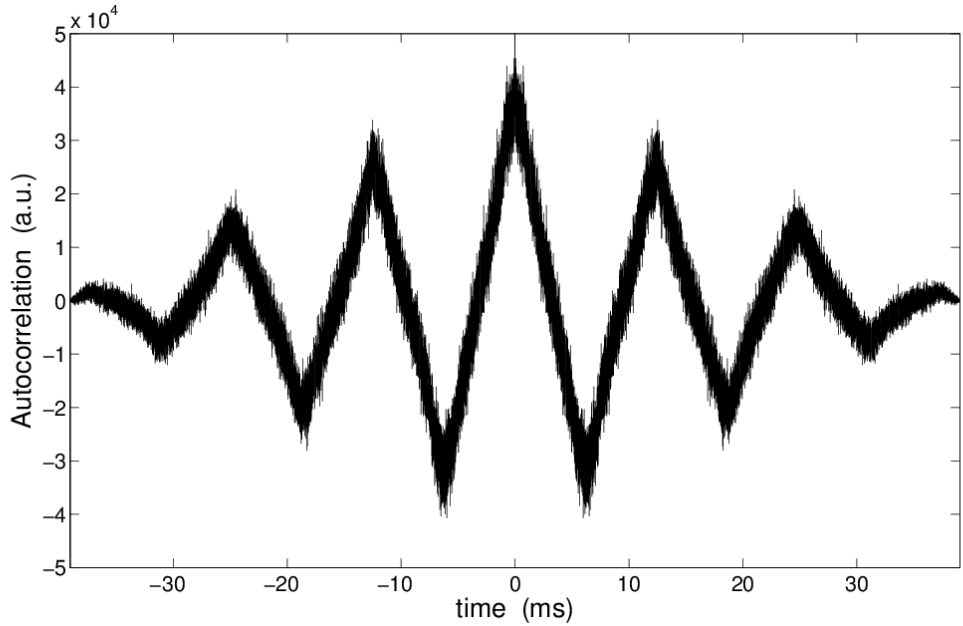

Figure 2

Autocorrelation (in arbitrary units) that uses the intensity as a function of time displayed in Figure 1 and Equation 1. The same 39 milliseconds of observations displayed in Figure 1 were used by the autocorrelation.

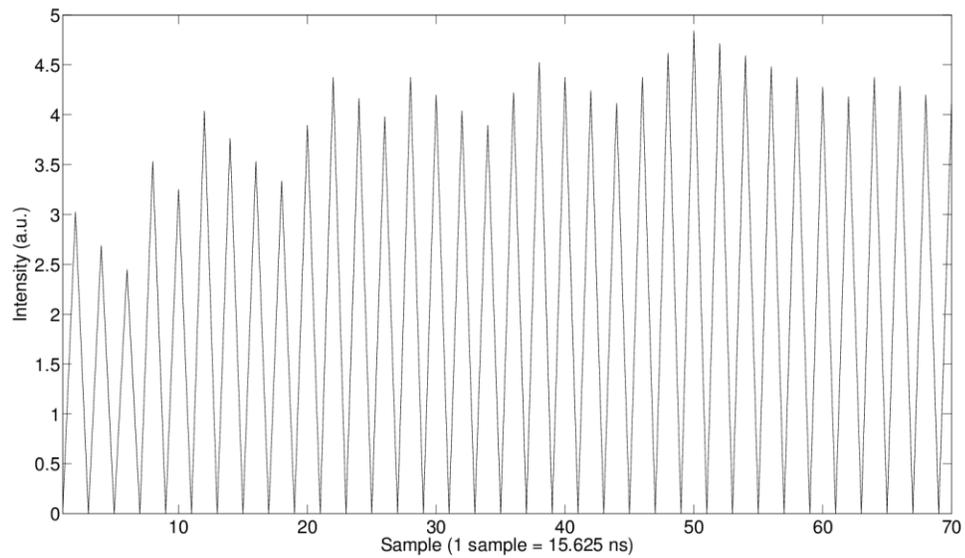

Figure 3

Intensity (in arbitrary units) as a function of sampling and time of a 5-MHz artificial signal produced by very narrow pulses injected on an exact 200 nanosecond period. The pulses seen in the figure are caused by the instrumentation sampling time (15.625 nanoseconds). The 5 MHz pulses are hidden in noise.

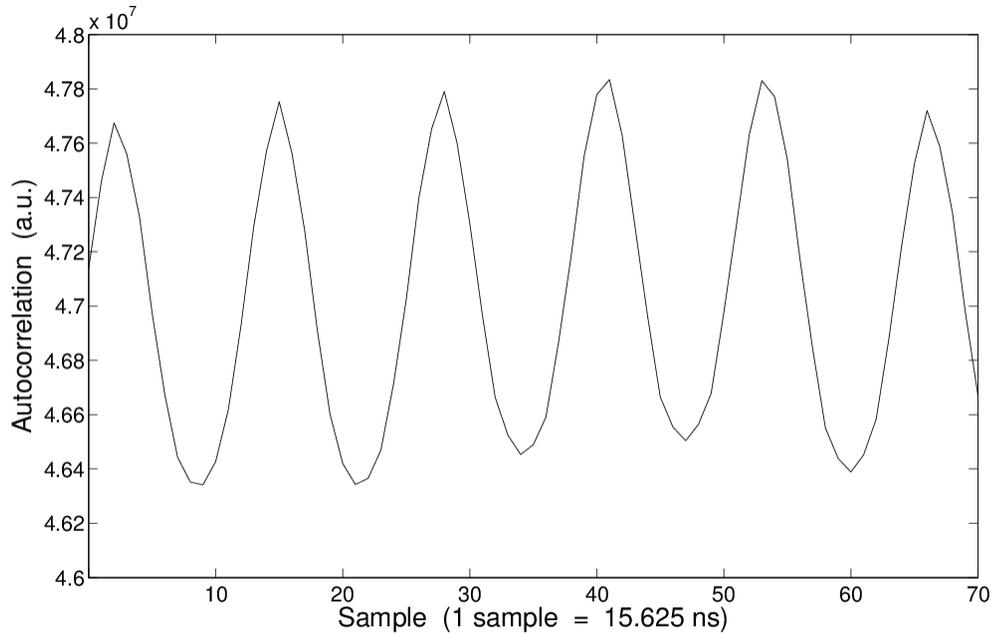

Figure 4

Autocorrelation (in arbitrary units) of the intensity as a function of time of an artificial signal that has a period of 200 nanoseconds (about 13 times the sampling time of 15.625 nanoseconds of the radio-astronomical data). We plot the intensity as a function of sampling number to clearly show that the autocorrelation can detect time variations comparable to the sampling time.

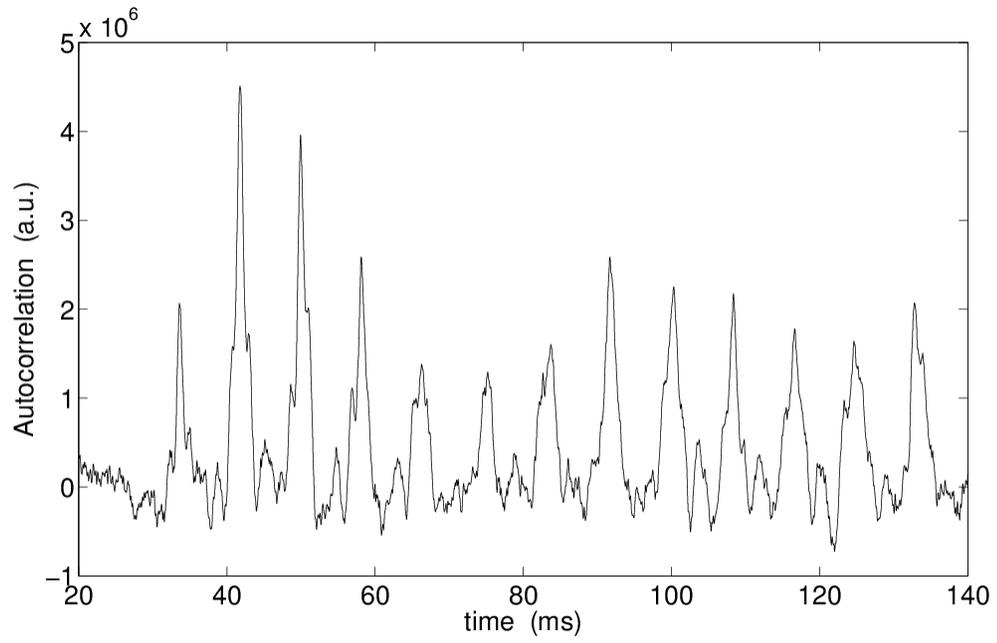

Figure 5

Autocorrelation (in arbitrary units) of the intensity as a function of time of an artificial quasi-periodic signal that has an average period of 8 milliseconds.

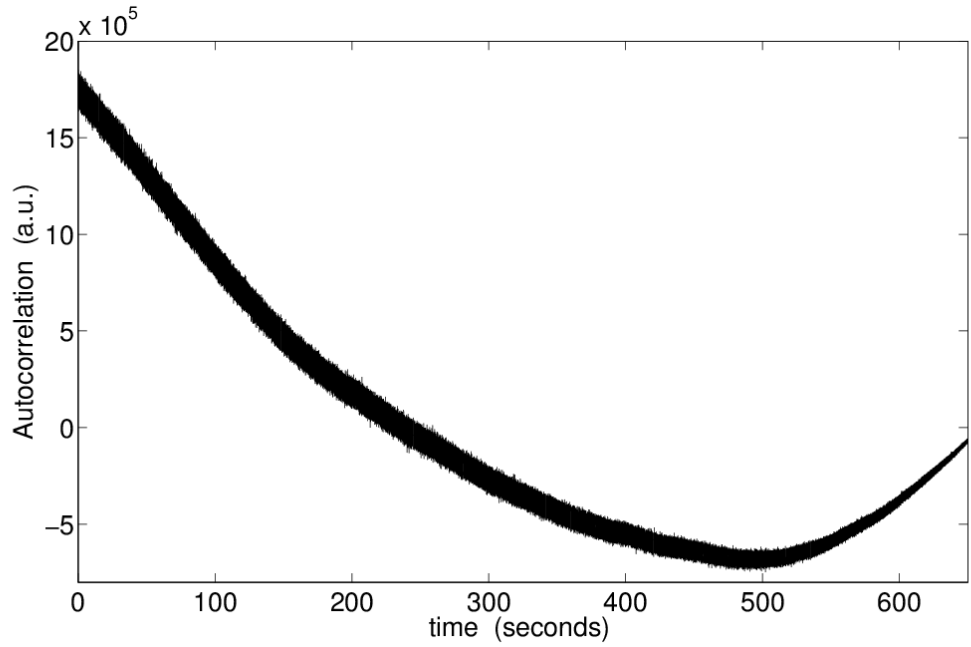

Figure 6

Autocorrelation (in arbitrary units) of intensity as a function of time containing non-periodic artificial signals. It uses the autocorrelation of 650 seconds of observations of the quasar PMN 0134-0931.